\documentclass[aps,prl,twocolumn,showpacs]{revtex4}
\usepackage{amssymb}
\usepackage{graphicx}
\usepackage{amsmath}
\usepackage{amsfonts}
\usepackage{bm}

\begin{document}

\title{Proposal for Creating a Spin-polarized $\bm{p_x+ip_y}$ State and Majorana Fermions}

\date{July 15, 2009}

\author {Patrick A. Lee}
\affiliation{Department of Physics, 
Massachusetts Institute of Technology, 
Cambridge, MA 02139} 

\begin{abstract}
The spin polarized $p_x + ip_y$ superconductor is known to support Majorana states but so far there has been no example.  We propose a simple setup to create such a state by depositing a ferromagnetic metal on a non-centrosymmetric superconductor with strong spin-orbit coupling which contains a Rashba-type term.
\end{abstract}
\maketitle

There has been a great deal of interest in producing the first example of Majorana fermions in the laboratory which obey non-abelian statistics.  This is an interesting phenomenon in its own right, but the excitement is compounded by the proposal that systems with non-abelian statistics can serve as the basis of topological quantum computation.\cite{Kitaev}  A great deal of attention has been focused on fractional quantum Hall states, notably at filling 5/2 where quasiparticles are predicted to be non-abelian.\cite{Moore,Nayak}  Recently, a novel proposal was made by Fu and Kane \cite{Fu} to deposit conventional $s$-wave superconductors on a topological insulator.  The vortices of the proximity-induced superconductor obey non-abelian statistics. Here we propose a simple geometry to create a spin-polarized triplet pairing of the $p_x+ip_y$ type.  This system is known to be equivalent to the 5/2 quantum Hall state.\cite{Read}  The vortices of this superconductor are understood to contain a zero energy Majorana state and obey non-abelian statistics.\cite{Read,Ivanov}   The advantage is that the energy scale can in principle be much higher than the quantum Hall state, making the creation and detection of the Majorana fermions easier to realize in the laboratory.

The present proposal is to deposit a thin ferromagnetic film on a superconductor with strong spin-orbit coupling and without inversion center.  There has been a great deal of experimental and theoretical activities on this type of non-centrosymmetric superconductors recently.\cite{Gorkov} Starting with the heavy fermion compound CePt$_3$S with superconducting $T_c$ of 0.7~K and its variants,\cite{Bauer,Frigeri} a number of non-centrosymmetric superconductors which are not heavy fermions have also been discovered.  For example, Li$_2$Pd$_3$B ($T_c$ = 6.7~K) and Li$_2$Pt$_3$B ($T_c$ = 2.43~K) have been studied.\cite{Yuan}  A more extensive list can be found in the paper by Yanase and Sigrist.\cite{Yanase}  Here we focus on CePt$_3$Si as an example which has been analyzed thoroughly.  As pointed out by Frigeri {\em et al.},\cite{Frigeri} the lack of inversion permits the term
\begin{equation}
H_p = \alpha \sum_{\bm k,\alpha ,\beta} \bm g_{\bm k} \cdot
\bm\sigma_{\alpha ,\beta} c^\dagger_{\bm k \alpha} c_{\bm k \beta}
\end{equation}
where $\bm g_{-\bm k} = -\bm g_{\bm k}$.  A familiar example of such term is the Rashba-type spin-orbit coupling, where
\begin{equation}
\bm g_k = (-k_y, k_x, 0)  .
\end{equation}
Once such asymmetric spin orbit coupling (ASOC) exists, the pairing symmetry can no longer be classified as singlet and triplet.\cite{Gorkov}  Instead, singlet pairing will induce triplet pairing.  Furthermore, Frigeri {\em et al.}\cite{Frigeri} showed that the most robust induced triplet pairing is of the form
\begin{equation}
\bm d(\bm k) \propto \bm g_{\bm k} ,
\end{equation}
where the pairing order parameter matrix is defined in the standard way as
\begin{equation}
\bm \Delta(\bm k) =
\left[
\begin{array}{cc}
\Delta_{++} & \Delta_{+-}   \\ 
\Delta_{-+} & \Delta _{--}  
\end{array}
\right]
=
\left[
\begin{array}{cc}
-d_x + id_y & d_z \\
d_z & d_x + id_y
\end{array}
\right]
\end{equation}
Combining Eqs.(2) and (3), we see that in the bulk, a triplet parallel spin pairing order parameter of the form ``$p_x \pm ip_y$'' exists for 
$\mp$ spin, respectively.
Now consider depositing a thin film of ferromagnetic metal on these superconductors.  An example is the strong ferromagnet (so called half metallic ferromagnet) CrO$_2$ where the spins are almost fully polarized.  In the bulk superconductor, spin is not a good quantum number due to spin-orbit coupling and $\pm$ in Eq.(4) should be understood as time reversed pseudo-spins.  On the other hand, the spin-orbit coupling in the ferromagnet is weak.  In the transmission of a quasiparticle between the superconductor and the ferromagnet, neither spin nor pseudo-spin is conserved but a + spin may preferentially convert to $\uparrow$ spin and vice versa.  Since only up-spin electrons are allowed in the ferromagnet, the proximity effect will induce only one component of the pairing amplitude $\Delta_{\uparrow\uparrow}$ in the ferromagnet, resulting in the spin-polarized $p_x - ip_y$ state that we are seeking.

The next question is how do we detect the existence of Majorana fermions once this setup is realized experimentally.  In addition to Majorana fermions bound to vortices, a chiral Majorana mode lives on the boundary of a finite dot of the ferromagnetic film.  A number of proposals have been made to detect Majorana fermions.\cite{Nilsson,Fu09,Akhmerov}  The simplest geometry is the one proposed by Law {\em et al.}\cite{Law} recently, where one simply tunnels into the Majorana edge state.  This can be achieved by moving a STM tip near the edge of the ferromagnetic dot 
deposited on the superconductor and measure the I-V characteristics.  At low enough temperatures it is predicted that the tunneling into the discrete Majorana levels (quantized by the finite circumference of the dot) is either zero or fully resonant with conductance of $e^2/h$, depending on whether there are an even or odd number of vortices on the dot.  This experiment requires a very simple geometry and gives unambiguous signatures of the Majorana mode.

It should be pointed out that this setup requires momentum parallel to the surface to be conserved in the transmission between the superconductor and the ferromagnet in order to preserve the momentum dependence of the pairing amplitude.  This requires high quality interface.  More generally, any scheme that relies on $p_x + ip_y$ pairing suffers from the fact that impurity scattering is pair breaking, so that high quality samples are required.  From this point of view, the proposal by Fu and Kane may have an advantage.
We also note that the CePt$_3$Si example given above is less than ideal in that $T_c$ is low and there is evidence that gap nodes exist in the bulk.\cite{Yanase}  Part of the purpose of this note is to motivate experimentalists to search for non-centrosymmetric superconductors which are not limited by the low $T_c$ of heavy fermions but which possess the requisite $\bm g_{\bm k}$ vector given in Eq.(2).  The analysis by Frigeri {\em et al.} leading to the Rashba form of Eq.(2) is quite general, and is applicable to any tetragonal structure which violates inversion symmetry and loses the mirror plane $(z \rightarrow -z)$.  Given the vast phase space of superconductors, one can be hopeful that such superconductors with a reasonably high $T_c$ can be found.
Furthermore, at an abrupt interface of MBE grown samples, a Rashba term is expected which may be sufficient for our purpose, even if we have a conventional centrosymmetric superconductor in the bulk.

After the completion of this work, a paper by Sau {\em et al.} \cite{Sau} appeared which proposed a mechanism for producing Majarano fermions using the proximity of a semiconductor with spin-orbit coupling to a superconductor and a ferromagnet.  Their proposal is closer in spirit to that of Fu and Kane but the physical idea of combining asymmetric spin orbit coupling, ferromagnetism and superconductivity is in common with the present work.  In particular, their pairing order parameter 
also contains an admixture of $p_x + ip_y$ states.

After the initial posting of this note, I received some important comments.  First, the idea of adding a spin-splitting term (in their case by an external Zeeman magnetic field) to non-centrosymmetric superconductors to produce a Majorana mode was discussed by Sato and Fujimoto.\cite{Sato}  They worked out in detail the parameter regime where nontrivial topological phases are to be expected.  Second, it was pointed out by C.L. Kane that unless the ferromagnetic layer in the present scheme is strictly two dimensional (a monolayer or a quantum well with a single sub-band), the existence of a Majorana mode is not guaranteed.  For example, an even number of quasi-two dimensional planes will produce an even number of Majorana fermions, which will admix to become ordinary fermions.  A variation of our scheme is to use a ferromagnetic insulator instead and grow it on top of a superconductor with strong orbit coupling.  Since the discontinuity at the interface will break inversion symmetry and produce the Rashba term, the superconductor does not need to break inversion symmetry in the bulk.  The quasiparticles will see the exchange field of the ferromagnet by proximity.  Under the right circumstances, a spin-polarized surface state may form at the interface which can support a $p + ip$ state.  We plan to perform some modelling to map out the parameter space for this to occur.

The support by NSF under DMR--0804040 is acknowledged.

\end{document}